\documentclass[reprint,aps,prx,floatfix,twocolumn]{revtex4-2}

\usepackage{amsfonts}
\usepackage{amssymb}
\usepackage{amsmath}
\usepackage{graphicx}
\usepackage{rotating}
\usepackage{epsfig}
\usepackage{psfrag}
\usepackage{bbm}
\usepackage{subfigure}
\usepackage{braket}
\usepackage{tikz}
\usepackage{stmaryrd}
\usepackage{wasysym}
\usepackage{array}
\usepackage{booktabs}
\usepackage{colortbl}
\usepackage[table]{xcolor}
\usepackage[unicode=true,colorlinks=true,pdfpagemode=UseOutlines,pdfstartview=Fith]{hyperref}
\hypersetup{linkcolor=blue,citecolor=blue,urlcolor=blue}

\begin{document}
\title{Error stabilized logical qubits in qudit generalizations of the monitored Kitaev model}

\author{Aayush Vijayvargia*, Ezra Day-Roberts*, Onur Erten}
\affiliation{Department of Physics, Arizona State University, Tempe, AZ 85287, USA}
\begin{abstract}
Monitored dynamics in quantum circuits provide tunable platforms for the realization of novel non-equilibrium phases. Motivated by recent advances in monitored Kitaev circuits, we investigate the monitored dynamics of the qudit ($d=4$) generalizations of the Kitaev model on the honeycomb and square lattices. In the absence of additional perturbations, the measurement-only dynamics of these models map onto multi-flavor loop models and display either critical or area-law entanglement scaling. Magnetic field terms couple different flavors and when measured with sufficiently large probability, they enhance the stability of the area-law phase that hosts the logical qubits. In a circuit picture, these terms correspond to single-qubit measurements and can be interpreted as errors. We also examine the impact of two-qubit measurements that commute with the plaquette operator, which induce effective non-quadratic interactions between Majorana fermions. These interactions can drive a transition to a volume-law-entangled phase and, for sufficiently strong coupling, stabilize a distinct area-law phase with an additional logical qubit for the square lattice model. Our results reveal a rich interplay between quantum spin liquids and monitored circuit dynamics, highlighting new mechanisms for engineering and controlling entanglement phases in multi-flavor Majorana systems.

\end{abstract}
\maketitle

\section{Introduction}
The advent of programmable quantum devices allows for precise control in shaping the entanglement structure of quantum many-body systems. A central setting is the study of monitored quantum circuits, where the competition between entangling unitary dynamics and disentangling projective measurements gives rise to a class of non-equilibrium phase transitions\cite{Skinner_PRX2019,Li2019,Gullans_PRX2020,Choi_PRL2020,Fisher_annrev2023,Zabalo_PRL2022}. Initially with the understanding in which measurements suppress quantum correlations, this view has been revised by the observation that measurement-only dynamics, without unitary evolution, can generate a variety of entanglement phases \cite{Lavasani_NatPhys2021,Sang_PRR2021,Klocke_PRB2022,Ippoliti_PRX2021,Lavasani_PRL2021,Hastings_Quant2021,Vu_PRL2024}. By tuning the probabilities or choosing different types of measurement operators, these systems can exhibit phases with local entanglement (area-law), extensive entanglement (volume-law), and intermediate critical phases characterized by multiplicative logarithmic corrections to the area-law. Symmetry\cite{Agrawal_PRX2022,Barratt_PRL2022} and topology\cite{Lavasani_NatPhys2021,Lavasani_PRL2021} further structure the phase diagram, yielding phases that can protect quantum information.

\begin{figure}[t!]    \includegraphics[width=1\linewidth]{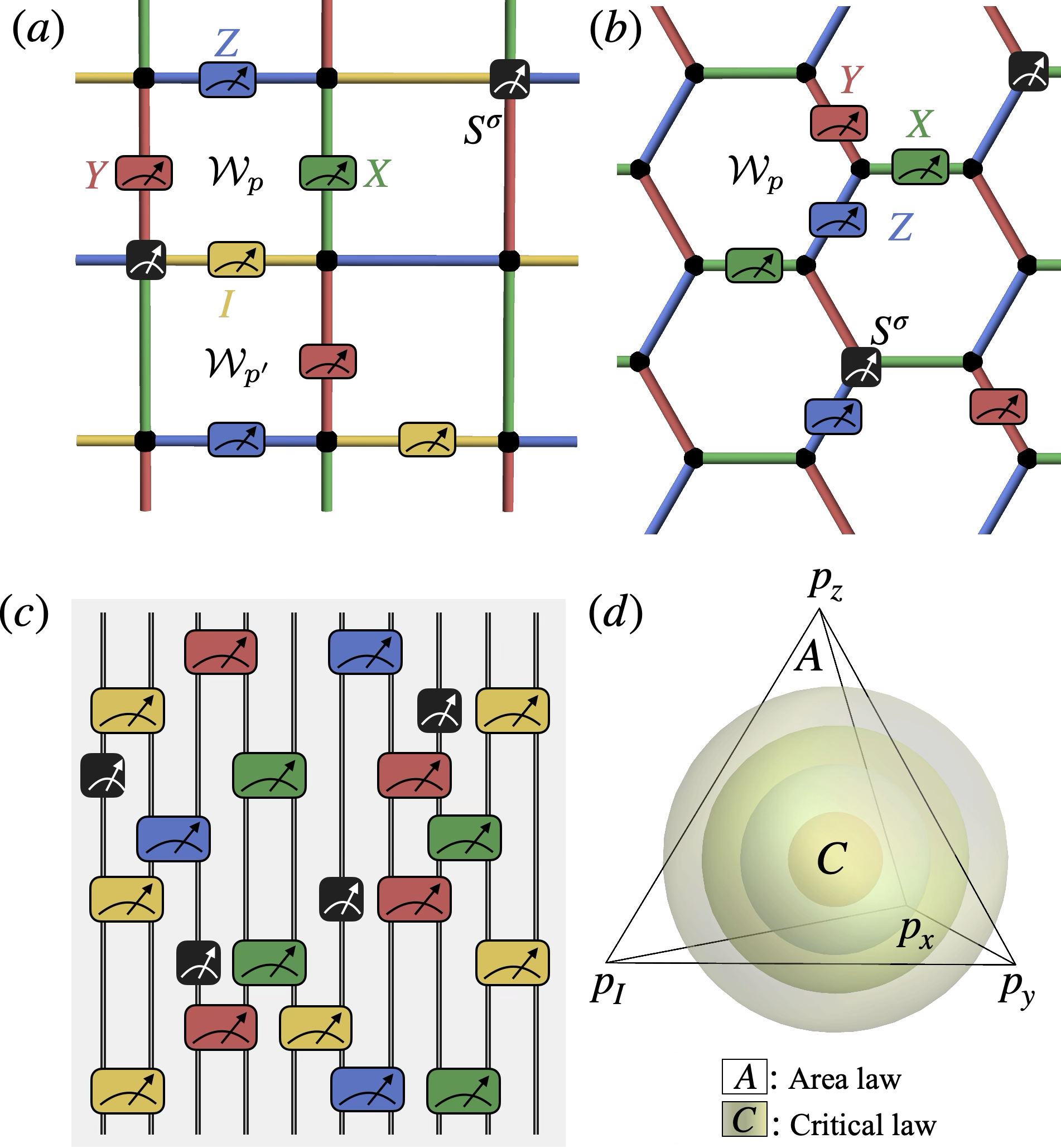}
    \caption{(a),(b) Schematic of our circuit models for square and honeycomb lattices. Two-qudit measurement checks act along bonds; for an $x-$bond a representative operator is $P_i^\sigma P_j^\sigma X_i^\tau X_j^\tau$. Single-site $S_i^\sigma$ measurements are indicated by black squares. (c) Circuit schematic: at discrete time steps, random single or two qubit checks are performed. (d) Square lattice phase diagram for $(p_x,p_y,p_z,p_I)$: tuning $p_h$ shifts the boundary between the critical and the area-law phase; beyond a threshold $p_h^\star$ the critical region vanishes and the entire tetrahedral exhibits area-law scaling with two logical qubits.}
    \label{Fig:1}
\end{figure}

This connection between monitored dynamics and error correction is most transparent in models that exhibit topological order, such as the toric code\cite{Kitaev_AnnPhys2003}, where the repeated measurements of commuting stabilizers protect a topological area-law phase against local errors which allows encoding of logical qubits within a larger set of qubits. A natural extension is to consider subsystem codes\cite{Bacon_PRA2006,Aliferis_PRL2007}, where the measurement operators, or `checks', do not mutually commute. The Kitaev honeycomb model provides a standard platform for exploring this physics\cite{Kitaev_AnnPhys2006}. Prior studies of its monitored dynamics show that a deterministic, time-periodic (Floquet) measurement schedule can dynamically generate logical qubits in an area-law phase \cite{Hastings_Quant2021,Gidney_Quantum2021faulttolerant,Haah_Quantum2022boundarieshoneycomb,Aasen_PRB2022,Paetznick_PRXQ2023,Zhu_arxiv2023}, while a stochastic protocol of random bond measurements yields a phase diagram featuring a topological area-law phase alongside a critical phase \cite{Lavasani_PRB2023,Sriram_PRB2023,Zhu_PRR2024}. The latter is understood as a gapless free Majorana liquid, where a Fermi surface produces the characteristic $L \ln L $ entanglement scaling\cite{Wolf_PRL2006,Gioev_PRL2006}, contrasting the Dirac dispersion observed in the corresponding phase of the Kitaev spin liquid. Furthermore, a highly entangled volume-law scaling phase can be realized within monitored Kitaev circuits, however, that involves measurement of long Pauli strings that can be understood as interacting Majorana liquid\cite{Zhu_PRR2024}.

In parallel, quantum information processing with qudits (generalization of a qubit to a $d$-level system) is beginning to become a practical alternative to the qubit framework. The larger local Hilbert space can enhance the efficiency of quantum algorithms\cite{Muthukrishnan_PRA2000,Bartlett_PRA2002,Zilic_IEEE2007,Gao_Quantum2023roleofentanglement,Nikolaeva_EPJQuant2024} and error-correction schemes \cite{Krishna_PRL2019,Watson_PRA2015}. Additional quantum levels can reduce circuit depth and provide new resources for encoding and manipulating quantum information, making qudit architectures relevant for current devices \cite{Andrist_PRA2015,deOliveira_NatCom2025,Kazmina_PRA2024,Islam_SciAdv2017,Weggemans_Quantum2022,Muminov_arxiv2025,romanova_arxiv2025,Kiktenko_RevModPhy2025}. Motivated by these considerations, we study the monitored dynamics of generalized Kitaev models built from four-dimensional qudits on both the square and honeycomb lattices\cite{Wu_PRB2009,Chulliparambil_PRB2020,Seifert_PRL2020,Chulliparambil_PRB2021,Vijayvargia_PRR2023,Akram_PRB2023,Nica_npjQM2023}. The enlarged on-site Hilbert space permits families of measurement operators—single-site Pauli operators and two-site interaction terms—that are local and compatible with the model’s gauge structure, as they commute with the conserved plaquette flux operators.

Motivated by these observations, we construct monitored qudit circuits on honeycomb and square lattices in which all checks are local and commute with plaquette fluxes. We use information-theoretic diagnostics, most notably the tripartite mutual information $I_3$ to explore the long-time steady-state phases. We then study the entanglement phases diagrams as a function of two control knobs: single-site measurements with rate $p_h$ and nearest-neighbor interaction measurements with rate $p_J$. The main findings are summarized below; detailed phase diagrams and finite-size analysis follow in the Results section.

\subsection{Summary of results}
We investigate the non-equilibrium steady states of monitored dynamics in qudit generalizations of the Kitaev model on both honeycomb and square lattices. The evolution is driven exclusively by local, plaquette-conserving projective measurements, providing a framework for realizing a rich entanglement phase diagram within a single, resource-efficient model. By tuning the relative probabilities of different measurement operators, the system exhibits topological area-law, critical, and volume-law entangled phases.

In the absence of additional perturbations, the competition between non-commuting bond measurements gives rise to two distinct regimes. When measurements of one bond type dominate, the system settles into a topologically ordered area-law phase. This phase functions as a dynamically generated quantum error-correcting code, protecting two logical qubits on a torus. Its entanglement structure is characterized by a bipartite entropy scaling as $S_A\approx \alpha L-S_{
\rm topo}$, where L is the length of the boundary and $S_{\rm topo}
 =1$ (in bits) is the topological entanglement entropy.  Conversely, when bond measurement probabilities are comparable, the system enters a critical phase with an entanglement entropy that logarithmically violates the area-law, scaling as $S_A\sim L\ln L$. This behavior is characteristic of a gapless free-fermion liquid with a Majorana Fermi surface. Unlike the area-law phase, this critical state provides no long-term quantum memory, as any initial state purifies to a product state on a polynomial timescale.

A key finding is the stabilizing role of single-site measurements, which in other contexts might be considered decohering errors. At finite measurement rates $p_h$, these local probes, which mix different flavors of Majorana fermions, enhance the stability of the topological area-law phase. Interestingly, this effect is non-monotonic: at small $p_h$, flavor-mixing can stitch together strings of different flavors, tending to favor long strings and transiently expanding the critical region. However, at higher rates, these measurements preferentially create short-range, localized stabilizers, strongly favoring the area-law phase. This stabilizing effect is particularly dramatic on the square lattice, where beyond a threshold rate $p_h^*$, the critical phase is completely suppressed, leaving a robust, topologically ordered state that spans the entire phase diagram as depicted in Fig.~\ref{Fig:1}(d). This demonstrates a powerful mechanism for stabilizing quantum information using only local monitoring.

Furthermore, we introduce two-qubit interaction measurements, which correspond to effective four-fermion interactions in the Majorana fermion representation and drive the system beyond a simple free-fermion description. These interactions can generate a volume-law entangled phase, with entropy scaling as $S_A\sim L^2$, using only local, nearest-neighbor resources. This circumvents the need for long-range operators or unitary scrambling typically invoked to create such extensive entanglement. The ultimate fate of the system under these interactions depends crucially on the underlying lattice geometry and interaction symmetry. On the honeycomb lattice, a Heisenberg-type interaction drives the system into a volume-law phase for any non-zero rate $p_J>0$. In contrast, the square lattice model, with its Ising-type interaction, exhibits a richer structure. It first transitions to a volume-law phase but at a sufficiently high interaction rate, undergoes a second transition into a distinct gapped area-law phase. This new phase is characterized by an increased tripartite mutual information of $I_3=3$, along with the emergence of a third logical qubit.

\section{Model and Methods}
\subsection{Qudit generalizations of the Kitaev model}
Before introducing the qudit generalizations of the Kitaev model, we first briefly overview the original Kitaev model defined on a honeycomb lattice\cite{Kitaev_AnnPhys2006}. The Hamiltonian has Ising type interactions that are distinct for three bonds and is given as $H_K = \sum_{\langle ij \rangle_\alpha} K_\alpha \sigma_i^\alpha \sigma_j^\alpha$ where $\alpha = x, y, z$ and $\sigma^\alpha$ are Pauli matrices. An important property of the Kitaev model is the presence of conserved flux operators, $\mathcal{W}_p$ defined around each plaquette.  These operators are formed by taking the product of bond operators along the edges of a plaquette. Importantly, the plaquette loop operators commute with each other $[\mathcal{W}_p, \mathcal{W}_{p' }]=0$ and with the Hamiltonian $[\mathcal{W}_p, H ]=0$. As a result, they are constants of motions, and they partition the Hilbert space into different sectors labeled by the flux eigenvalues, $\pm1$. The exact solution can be obtained by introducing four Majorana fermions per site, $\sigma_i^\alpha = i b_i^\alpha c_i$ which lead to 
\begin{eqnarray}
H_K &=& \sum_{\langle ij\rangle_{\alpha}}-K_\alpha iu_{ij}^{\alpha}c_ic_j
\end{eqnarray}
where $u_{ij}^\alpha=ib_{i}^{\alpha}b_{j}^{\alpha}$. The bond operators $u_{ij}^\alpha$ also commute with the Hamiltonian, $[H, u_{ij}^\alpha] = 0$, and each flux operator can be expressed as a product of the bond operator around the plaquette.

While theoretically quite elegant, the topologically ordered ground state in the Kitaev model can be easily destroyed upon the application of external magnetic field or additional interactions since these terms do not commute with the flux operator\cite{Rau_PRL2014}. One way to fix this problem is to expand the local Hilbert space. The original Kitaev model is defined for $s=1/2$ degrees of freedom (DOF) with a local $d=2$ dimensional Hilbert space, corresponding to a qubit DOF per site. However, it is possible to extend Kitaev's construction to higher dimensions, $d=2^n$, using Clifford algebra\cite{Wu_PRB2009}. These generalizations of the Kitaev model have qudit DOF per site. In a condensed matter setting, the additional DOF can correspond to orbital DOF and therefore these models are commonly referred as the Kitaev spin-orbital models\cite{Wu_PRB2009, Yao_PRL2011, Nakai_PRB2012, Carvalho_PRB2018, Chulliparambil_PRB2020, Natori_PRL2020, Chulliparambil_PRB2021, Seifert_PRL2020, Tsvelik_PRB2022, Nica_npjQM2023, Akram_PRB2023, Vijayvargia_PRR2023, Keskiner_PRB2023, Poliakov_PRB2024, Majumder_PRB2024, Dutta_arXiv2025, Neehus_arXiv2025, Keskiner_MatTodayQ}. In this article, we consider qudit generalizations with four-dimensional ($d=4$) local Hilbert space defined on honeycomb and square lattices. The Hamiltonians for these two models are 
\begin{eqnarray}
        H_{YL} &=& \sum_{\langle ij\rangle_{\alpha}} K_\alpha(\boldsymbol{\sigma}_i \cdot \boldsymbol{\sigma}_j)\otimes\tau_{i}^{\alpha}\tau_{j}^{\alpha} \label{eq:YL}\\
    H_{SqL} &=& \sum_{\langle ij\rangle_{\alpha}} K_\alpha(\sigma_{i}^{x}\sigma_{j}^{x}+\sigma_{i}^{y}\sigma_{j}^{y})\otimes\tau_{i}^{\alpha}\tau_{j}^{\alpha} \label{eq:SqL}
\end{eqnarray}
$H_{YL}$ is the Yao-Lee model\cite{Yao_PRL2011} defined on a honeycomb lattice where $\alpha = x, y, z$ as schematically depicted in Fig.~\ref{Fig:1}. $H_{SqL}$ is the square lattice (SqL) generalization of the Kitaev model\cite{Nakai_PRB2012}. In this case the four bonds are labeled as $\alpha = x, y, z, I$ where $I$ corresponds to the identity operator on the orbital ($\tau$) sector (see Fig. \ref{Fig:1}(a)). The $\sigma$ and $\tau$ DOF correspond to two independent sets of Pauli matrices. Similar to Kitaev model, there are plaquette operators for both models that commute with the Hamiltonians. For the YL model, they are given as $\mathcal{W}_p^{YL}=\mathbbm{1}\otimes\tau_i^x\tau_j^y\tau_k^z\tau_l^x\tau_m^y\tau_n^z$ (see Fig.~\ref{Fig:1}(b)). For the SqL model the bond-dependent exchange interactions double the unit cell resulting in two distinct plaquettes, $\mathcal{W}_p^{SqL}=\sigma_k^z\sigma_n^z\otimes\tau_i^z\tau_j^x\tau_k^z\tau_n^x$, and $\mathcal{W}_{p'}^{SqL}=\sigma_k^z\sigma_n^z\otimes\tau_k^x\tau_l^z\tau_m^x \tau_n^z$ as shown in Fig \ref{Fig:1}(a).

The exact solutions of $H_{YL}$ and $H_{SqL}$ can be obtained by representing $\sigma$ and $\tau$ operators in terms of four-dimensional $\Gamma$ matrices that obey the Clifford algebra and introducing six Majorana fermions per site (see Appendix~\ref{Appdx:qudit} for details),
\begin{eqnarray}
H_{YL} &=& \sum_{\langle ij\rangle_{\alpha}}-K_\alpha iu_{ij}^{\alpha}(c^x_ic^x_j+c^y_ic^y_j+c^z_ic^z_j) \label{eq:MFYL}\\
H_{SqL} &=& \sum_{\langle ij\rangle_{\alpha}}-K_\alpha iu_{ij}^{\alpha}(c^x_ic^x_j+c^y_ic^y_j) \label{eq:MFSqL}
\end{eqnarray}
where $u_{ij}^\alpha=ib_{i}^{\alpha}b_{j}^{\alpha}$. Similar to the Kitaev model, the free Majorana fermions move in the background of $\mathbb{Z}_2$ gauge fields. However, in these cases there are 2 and 3 flavors of free Majorana fermions for SqL and YL models respectively. 

The Majorana representation introduces an overcomplete Hilbert space. Therefore, the physical states must satisfy the constraint $D_{i}=ib^1_{i}b^2_{i}b^3_{i}b^4_{i}b^5_{i}c$. This constraint can be implemented by a projection operator, $P=\prod P_i$ where $P_i = (1+D_i)/2$. The physical wave function can be obtained by applying the projection operator on a gauge-fixed wave function, $\Psi_{\rm Phys} = P \Psi_f$.

As mentioned above, there are certain perturbations that commute with the plaquette operators for YL and SqL models. In particular all three $\sigma_i^\alpha$ operators commute with $\mathcal{W}_p^{YL}$ and $\sigma_i^z$ commutes with $\mathcal{W}_{p/p'}^{SqL}$. Therefore, it is possible to include a magnetic field term to these models while preserving the exact solvability,
$\sum_i \bf{h_i} \cdot \boldsymbol{\sigma}_i$ for YL model and $h_i^z \sigma_i^z$ for SqL model. These terms can shift the ground state flux sector and give rise to a Majorana Fermi surface\cite{Chulliparambil_PRB2021}. In addition, it is possible to include exchange terms involving $\sigma$ DOF that commute with the flux operators, $J\sum_{\langle ij \rangle} \boldsymbol{\sigma}_i \cdot \boldsymbol{\sigma}_j$ for the YL model and $J\sum_{\langle ij \rangle} \sigma^z_i \sigma^z_j$ for the SqL model. These exchange terms can give rise to magnetic order in $\sigma$ DOF while preserving a liquid character in $\tau$ DOF\cite{Seifert_PRL2020, Akram_arXiv2025_2}. We will utilize these $h$ and $J$ terms to tune the entanglement phase diagrams of the measurement-only circuits. Note that for the original Kitaev model, it is not possible to include similar terms that commute with the flux operators.

\begin{figure*}
    \centering
    \includegraphics[width=0.9\linewidth]{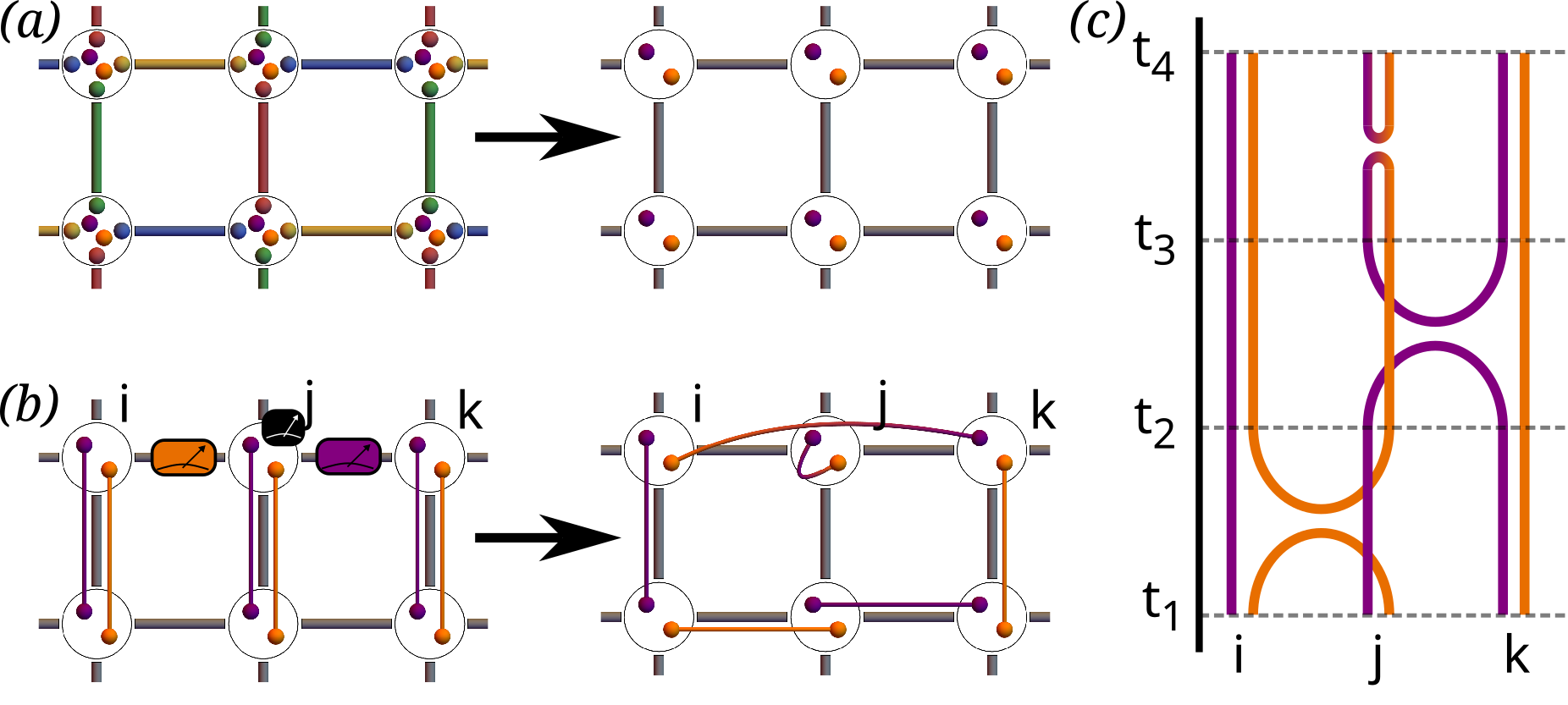}
    \caption{(a) Measurement of plaquettes leads to flux purification where the dynamics of the density matrix only depends on the dynamics of the Majorana fermions. There are two flavors of Majorana fermions for the square lattice model represented by purple and orange circles. (b) Example measurements of both species of Majorana fermions as well as onsite $\sigma_z$ term, marked as boxes, and the resulting string configuration. (c) Worldlines corresponding to measurements in (b) for three labeled sites.}
    \label{Fig:2}
\end{figure*}

\subsection{Circuit dynamics}
Next, we construct the monitored circuit dynamics for the generalized Kitaev models. The core idea is to interpret the terms in the Hamiltonians as measurement operators. This perspective frames the dynamics within the language of subsystem codes, where the nearest-neighbor bond terms act as parity checks \cite{Bacon_PRA2006,Sriram_PRB2023}. Unlike a stabilizer code, these
checks are allowed to anti-commute, however they can be measured sequentially to measure stabilizers of the group that commute with further checks. In our model, the product of these checks around any elementary plaquette forms the plaquette operator $\mathcal{W}_p$. As previously established, these operators commute with all individual bond operators. Consequently, once the system is initialized in a specific flux sector, 
it remains there throughout the measurement-only evolution. 

Our simulations are performed on a periodic $L \times L$ lattice (a torus) each hosting a four-dimensional qudit with $N=L^2$ and $N=2L^2$ sites for square and honeycomb lattices respectively. The dynamics proceed in discrete time steps, where each step consists of $N$ projective measurements (see Fig.~\ref{Fig:1}(c)). For each measurement, an operator is chosen stochastically and measured. The protocol for a single measurement involves first selecting an operator type based on prescribed probabilities: with probability $p_\alpha$, where $\alpha \in \{x, y, z, I\}$ (no $I$- bond for the honeycomb lattice model), a bond of type $\alpha$ is selected. We also allow for additional plaquette conserving operators: with probability $p_h$, a single-site, plaquette-conserving operator in the $\sigma$-sector is chosen or with probability $p_J$, a nearest-neighbor interaction term is selected (Ising-type $Z^\sigma_i Z^\sigma_j$ on the square lattice and Heisenberg-type $\vec{\sigma}_i \cdot \vec{\sigma}_j$ on the honeycomb lattice). After the measurement operator is chosen, a specific instance is selected uniformly at random from all available operators of that type. This operator is then projectively measured. Specifically, a bond measurement of type $\alpha$ between sites $i$ and $j$, the operator is not fixed but is instead drawn from an ensemble. For example, we perform the $z$-bond check by measuring the operator $\mathcal{O}^z_{ij} = P^\sigma_i P^\sigma_j Z^\tau_i Z^\tau_j$, where the Pauli operator $P \in \{X, Y, Z\}$ (only $P \in \{X,Y\}$ for square lattice model) is chosen uniformly at random for each individual measurement. This measurement schedule ensures that, on average, each qudit is involved in a $\mathcal{O}(1)$ number of measurements per time step, while preserving the initial flux configuration defined by the $\mathcal{W}_p$ operators.

Since all measurement operators are Pauli strings, the system's evolution constitutes a Clifford circuit. This allows for efficient classical simulation using the stabilizer formalism\cite{Gottesman_PRA2004}. The quantum state $\rho$ can be written in terms of the stabilizer generators: $\rho=\prod_{i=1}^N (\frac{1+g_i}{2})$. Each projective measurement is implemented by updating this tableau: if the measured operator commutes with the entire stabilizer group and is independent of it, it is added as a new generator; if it anti-commutes with some generators, it replaces one of them and the others are replaced by a commuting product of those generators. The conserved plaquette operators $\mathcal{W}_p$ remain fixed in the stabilizer group by construction. We investigate the system's properties by averaging over many stochastic quantum trajectories, each corresponding to a unique sequence of measurement locations, types, and outcomes. A significant advantage of the Clifford formalism is that the bipartite entanglement entropy of the state can be computed directly from the stabilizer group and is independent of the specific measurement outcomes (i.e., the eigenvalues). 
To expedite steady-state characterization, we initialize with all plaquette stabilizers projected onto an unentangled product state,
$\rho_{\mathrm{in}} = \Big(\prod_p \frac{1+\mathcal{W}_p}{2}\Big)\ket{0}^{\otimes N}$.
We then evolve under the prescribed measurement-only dynamics until the entanglement diagnostics reach a steady state, with a maximum of 1000 time steps.

\subsection{Multi-flavor loop models}
The entanglement properties of monitored circuit dynamics of free Majorana models can be captured by loop representations\cite{Nahum_PRR2020, Merritt_PRB2023, Klocke_PRX2023, Lavasani_PRB2023, Klocke_PRB2025}. Since our simulations do not rely on loop models, we omit the details of the mapping procedure and instead present the basic formulation, emphasizing the new features that emerge in the qudit generalizations.

In measurement-only circuit dynamics, the wave function evolves via local projective measurements. Considering a maximally mixed state to initialize the circuit simulations, the plaquette operators ($\mathcal{W}_p$),  are measured at a relatively short time, $t\sim O(\log L)$ as depicted in Fig.~\ref{Fig:2}(a)\cite{Lavasani_PRB2023}. Once $\mathcal{W}_p$'s are included in the stabilizer tableau, they cannot be removed as discussed in the previous section. This process is commonly referred as flux purification\cite{Klocke_PRB2025}. Once all fluxes are fixed, the full density matrix can be expressed as a product of the density matrix of the Majorana fermions and the fluxes, $\rho_f(t)\simeq \rho_c(t) \otimes \rho_W$. Note that the flux DOF don't have dynamics after flux purification and the whole dynamics can be captured by the Majorana fermions. 

In order to illustrate the measurement-only dynamics of Majorana fermions, consider a state for the square lattice model where the Majorana fermions are paired along the $z$ direction between neighboring sites as depicted in Fig.~\ref{Fig:2}(b) (left panel). Note that the square lattice model has two flavor of itinerant Majorana fermions, shown by purple and orange colors for $x$ and $y$ flavors respectively. Next, consider a measurement between $i$ and $j$ sites which are connected by the $I$ (identity) bond, $\sigma_i^y\sigma_j^y$ with an outcome +1. This measurement can be implemented via a projector $(1+c_i^xc_j^x)/2$. Application of this on the wave function results in the entangling of the $x$-flavor of Majorana fermions at sites $i$ and $j$ as well as the entangling of the lower sites at the other end of the preexisting strings. As a result the new stabilizers are included in the tableau. As such, it is possible to represent the Pauli strings in the stabilizer tableau as strings connecting pairs of sites. An example of how the strings change as a consequence of measuring two bond operators and an onsite term is shown in Fig.~\ref{Fig:2}(b) (right panel). The time evolution of the circuit represents a third dimension and the creation and annihilation of these strings form loops as depicted in Fig.~\ref{Fig:2}(c). Therefore, the dynamics of free Majorana fermion models can be captured by loop models in one higher dimension.

Loop models can be simulated via large-scale numerical methods. The simulations show that there are two phases: an area-law phase where the strings are short-ranged and a critial phase with long-range strings\cite{Nahum_PRR2020, Merritt_PRB2023, Klocke_PRX2023, Lavasani_PRB2023, Klocke_PRB2025}. 
In the qudit generalizations of the Kitaev model, the world lines of different flavors strings are decoupled from each other since the measurement operators of different flavors commute with each other. Therefore, the qudit generalizations of the Kitaev model map to multi-flavor loop models.

While the different flavors of loops are decoupled for the original models, the onsite $h-$term, $\sigma_i^\alpha = i\epsilon^{\alpha \beta\gamma} c_i^\beta c_i^\gamma/2$ hybridizes different flavors (Fig.~\ref{Fig:2}(b, c)). This flavor-mixing can create both longer strings by patching shorter string of different flavors as well as single site stabilizers (strings with $l=0$). As discussed in the Results section, our simulations indicate that the former effect is dominant for small $p_h$ and the latter is important for large $p_h$. 

We note that the two-site interaction term, measured with probability $p_J$, $\sigma_i^z\sigma_j^z = -c_i^xc_i^yc_j^xc_j^y$ does not simplify to bilinears in Majorana fermions. Therefore, it creates interactions among Majorana fermions and the model can no longer be mapped to a loop model. This interacting regime correlates with the emergence of the volume-law phase observed in our numerics (see Results). This volume-law phase cannot be realized within the Majorana loop models \cite{Klocke_PRB2025,Klocke_PRX2023}.

\begin{figure*}
    \centering
    \includegraphics[width=1\linewidth]{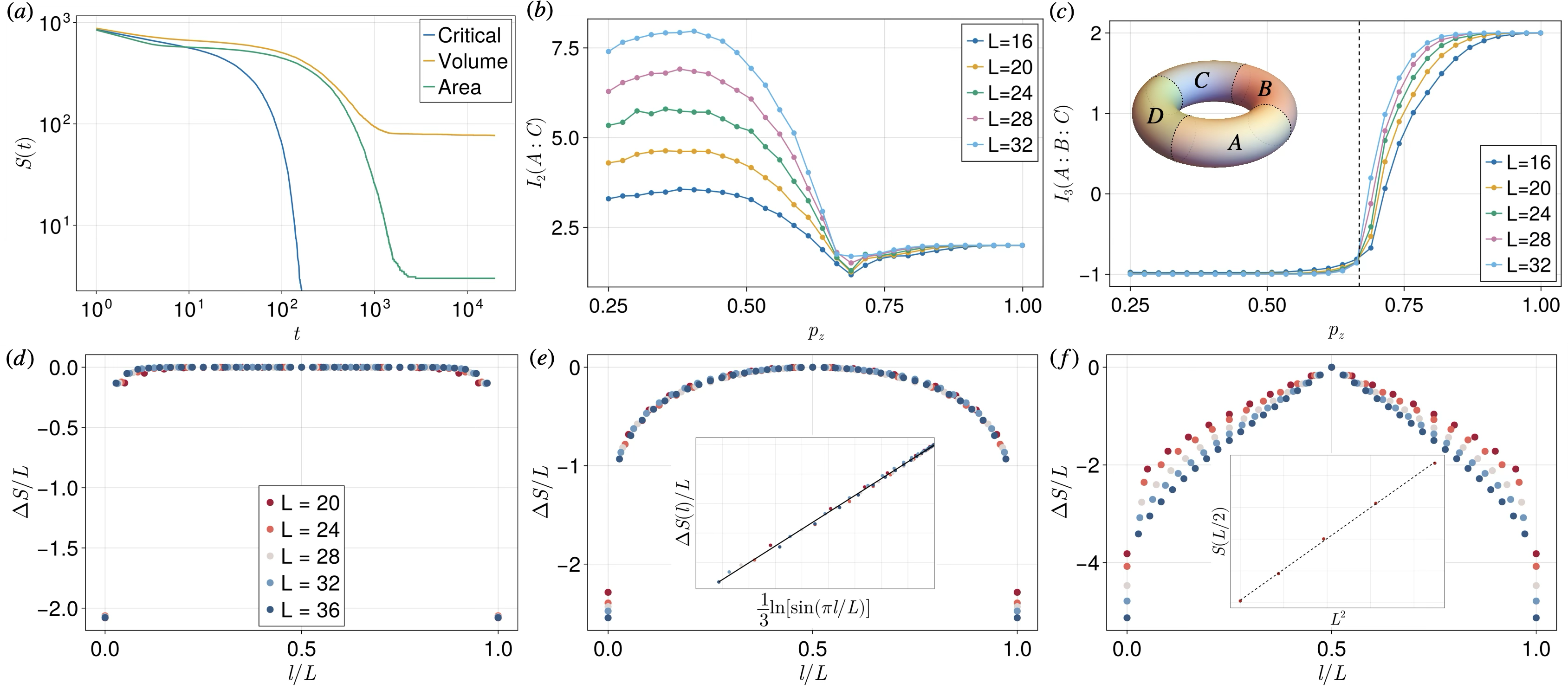} 
    \caption{(a)Dynamical purification from a fully mixed state for selected points in each of the three phases. All phases begin with an exponential decrease in entropy as the plaquettes are measured. After that, the critical phase quickly removes all remaining entropy. The area-law phase plateaus then descends to a constant value. The volume-law phase is similar but with a system size dependent final value. (b), (c) Bipartite and tripartite mutual information establishes a critical to area-law transition where curves cross as a function of $p_z$. (c) inset shows partitioning of the torus in four equal cylinders for mutual information calculations. Transition is marked by dashed line. (d, e, f) Subsystem entanglement entropy with cylinders of varying width: (d) In area-law (e) critical and (f) volume-law phases. Inset (e) data collapse of scaled entropy. Inset (f) shows data collapse of half system entanglement entropy showing $L^2$ scaling. All the data presented in the above panels is obtained for the square lattice model. The appropriate choice of parameters is made to display the representative properties of the respective phases.}
    \label{fig:purification}
\end{figure*}

\section{Results}

We now present the numerical analysis of monitored qudit circuits on honeycomb and square lattices. Phase boundaries are identified primarily from size-independent crossings of the tripartite mutual information $I_3(A:B:C)$ with finite-size collapse while subsystem entanglement and dynamical purification provide cross-checks. To compare probability ensembles on equal footing, we parametrize by the distance $r$ from the isotropic point inside the simplex (triangle for honeycomb, tetrahedron for square lattices) and report the normalized critical radius $r_c/r_{\rm edge}$, where $r_{\rm edge}$ is the distance to the edge center of the triangle or tetrahedron. 
We begin by discussing the entanglement diagnostics and then quantify how single-site measurements $p_h$ reshape the critical surface (Fig. \ref{fig:ph}), and then study the effect of two-site interaction measurements 
$p_J$ (Figs. \ref{fig:pj_cuts}-\ref{fig:pj_rc}).
\subsection{Entanglement measures}
The entanglement scaling for different phases follows an area-law, volume-law and a critical phase with $L$, $L^2$ and $L\log{L}$ scaling respectively. In order to distinguish these phases we use four entanglement diagnostics including dynamical purification, subsystem entanglement entropy, bipartite and tripartite mutual information. 


\subsubsection{Dynamical purification}
Dynamical purification starts with a maximally mixed state $\rho = I/4^N$ and time evolves this according to our monitored dynamics. The entropy of the state decays towards a pure state with zero entropy, and the speed of this is related to the entanglement phase, as shown in Fig. \ref{fig:purification}(a). Using the stabilizer formalism given above, it is easy to track the entropy of a state. For a state with $k$ generators $\rho = \prod_{i=1}^k (1+g_i)/2$ the entropy is simply $N-k$. 

In all states the plaquette DOF are first to purify, corresponding to the roughly linear portion in Fig. \ref{fig:purification}(a) and taking a polynomial length of time. The phases can be differentiated by the next portion, consisting of the purification of the remaining Majorana fermion DOF. In the critical phase, the state completes its evolution into a zero entropy pure state with the measurement of all DOF in polynomial timescale. However, in the area and volume-law there are remaining DOF that are not measured for exponential time scales\cite{Gullans_PRX2020,Ippoliti_PRX2021}. In the area-law, these correspond to the logical qubits protected by the subsystem code. For this choice of parameters, there are 2 remaining logical qubits corresponding to the two non-contractible loops spanning the two dimensions of the lattice. Notably, we also realize a distinct area-law-II phase with 3 logical qubits. However, determining the full purification requires many measurements for each system and it only gives a qualitative determination of the phase.

\subsubsection{Subsystem Entanglement Entropy}
The subsystem entanglement entropy scales with the subsystem size differently in the different phases. Fig. \ref{fig:purification}(d-f) show the normalized entanglement entropy of cylinders of size $\ell \times L$, relative to half system size, i.e. an $L/2 \times L$ cylinder, as a function of $\ell$. In the area-law phase long strings are exponentially suppressed, so the entropy is only significantly different from its half-system value for regions of size $\sim \ln L$. In the critical law phase there is a power law scaling of loop length leading to a divergence of the entropy as $S\sim L\ln L$. This also changes the entropy from its half-system value for a much larger range of cylinder size. There is a data collapse of the scaled entropy with a functional form \cite{Calabrese2004}
\begin{equation}
    S/L = a(L) + b(L) L \ln \left[\sin(\pi l/L)\right]
\end{equation}
as shown \ref{fig:purification}(e) inset.

The volume-law phase shows linear scaling for intermediate system sizes $l\sim L/4$, as well a cusp near $L/2$ known from Page scaling \cite{Page1993}. It also shows a clear failure of data collapse for different $L$, with the volume-law nature emphasized by the inset showing $L^2$ scaling of the half-system entropy.

The entanglement entropy allows for a quantitative measure of subleading contributions from other phases by fitting a combination of functional forms, which can be significant especially near phase boundaries. However, this property also makes it less convenient to determine exact phase boundaries, as differentiating between a very small contribution and zero contribution in an imperfect fit can be challenging.

\subsubsection{Bipartite mutual information}
The bipartite mutual information between regions $A$ and $B$ is defined as
\begin{equation}
    I_2(A:B) = S_A + S_B - S_{AB}
\end{equation}
and measures the information that is contained in the join region $AB$ but not in either region by itself. In terms of stabilizers, this counts the number of Pauli strings spanned by the generators that have support in $A$ or $B$ but not in the complement region $\overline{AB}$. In terms of the loop models, this counts the number of open strings with one end in $A$ and the other in $B$. We use a geometry dividing our system into four equal size cylinders, as shown in the inset of Fig. \ref{fig:purification}(c). 

Fig. \ref{fig:purification}(b) shows this quantity as a function of $p_z$ for the square lattice model with no additional terms along the $p_x=p_y$ line which shows a transition from a system size dependent $I_2(A:C)$ in the critical phase to a constant value in the area-law phase.

Similar to subsystem entanglement entropy, this measure does not provide a simple quantitative description of the phase boundary, especially given the non-monotonic behavior evident for smaller $L$. In addition, it does not allow simple differentiation between critical and volume-law phases.

\subsubsection{Tripartite mutual information}
The tripartite mutual information, $I_3(A:B:C)$ is defined by the inclusion-exclusion principle
\begin{align}
    I_3(A:B:C) =& \ I_2(A:B) + I_2(A:C) - I_2(A:BC) \\
    =& \ S_A + S_B + S_C - S_{AB} - S_{AC} - S_{BC}\nonumber\\
    &+ S_{ABC}
\end{align}
where we consider the same partition scheme described in the bipartite mutual information section. Tripartite mutual information subtracts out even more types of information, being concerned with information that is stored in the entirety of the region $ABC$ but not in any single or two region subset. 

We can quantitatively determine the transition point by using finite-size scaling. Although we do not produce a mapping to a system with a well-defined Landau-type phase transition, we observe a data collapse with the scaling form $I_3 = F[(r-r_c) L^{1/\nu}]$ for some transition radius $r_c$, corresponding to the crossing point of the curves, and scaling exponent $\nu$ similar to scaling observed in monitored Kitaev model \cite{Lavasani_PRB2023,Klocke_PRB2025}. 
The tripartite mutual information shows extensive behavior in the volume-law phase, as seen in Fig. \ref{fig:pj_cuts} and we can again use the crossing point to quantitatively determine the transition point. 

For these reasons, in the rest of the paper we use the tripartite mutual information to map our phase diagrams while benchmarking our results with the other entanglement measures for certain parameters.

\subsection{Entanglement phase diagrams}
\begin{figure}
    \centering
    \includegraphics[width=0.9\linewidth]{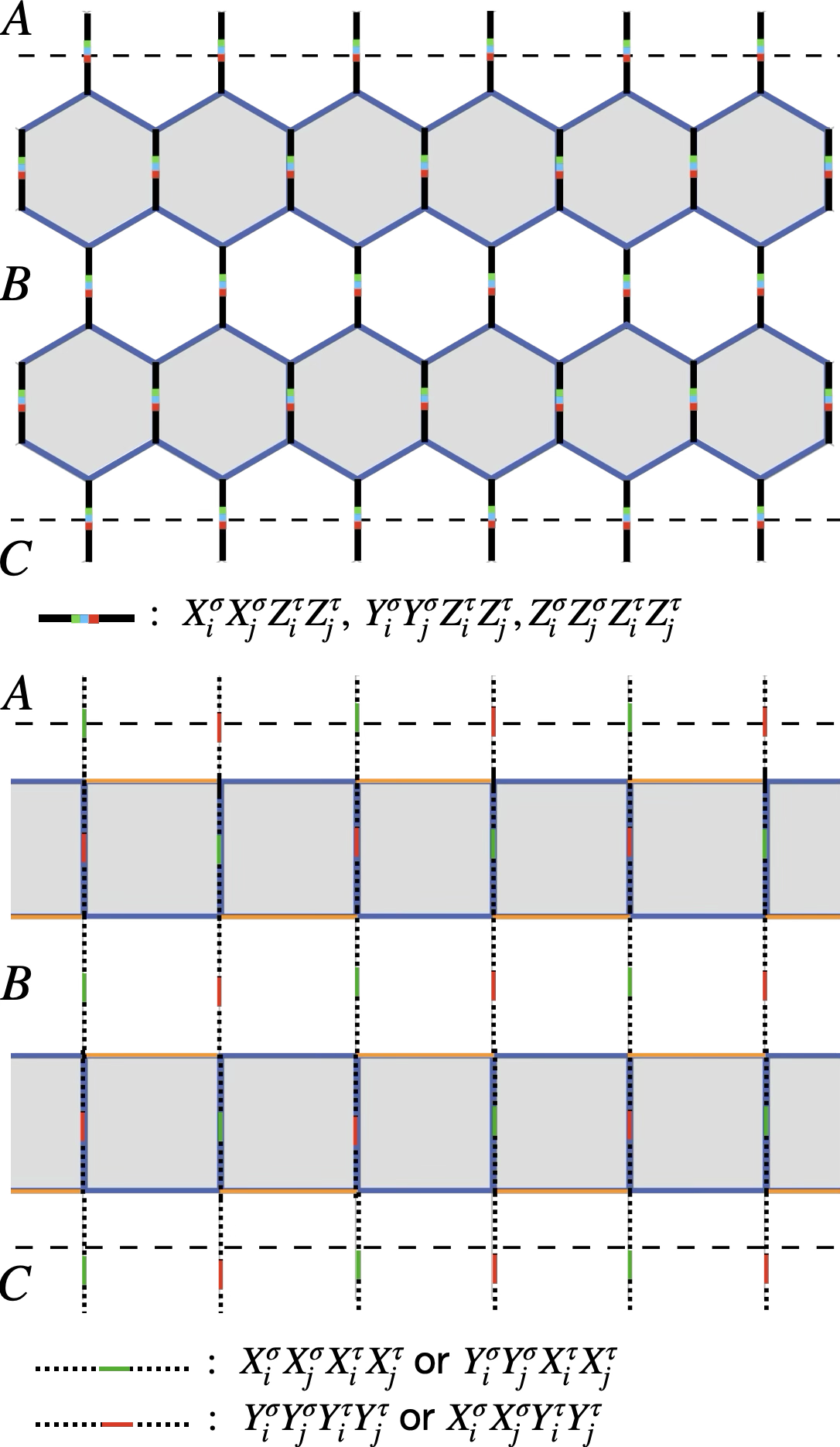}
    \caption{Stabilizers shared between regions $A$ and $C$ with trivial support on $B$ that contribute to the mutual information in the area-law phases for the honeycomb and square lattices respectively.}
    \label{Fig:4}
\end{figure}

\begin{figure}
    \centering
    \includegraphics[width=\linewidth]{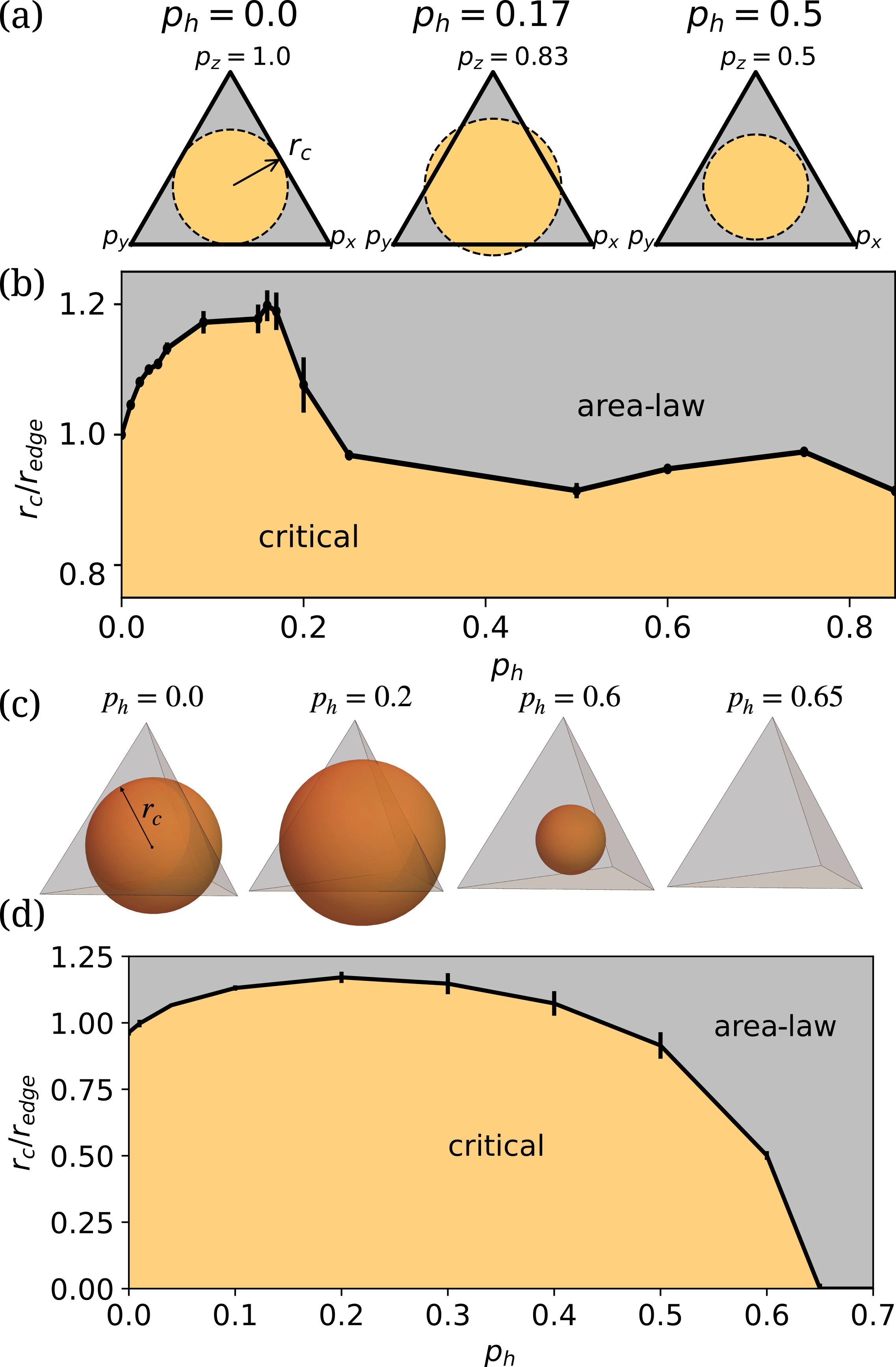} 
    \caption{Effect of single site term (a) critical phase (orange) and area-law (gray) in $p_x,p_y,p_z$ triangle for selected values of $p_h$ (b) phase diagram of radius vs $p_h$ in honeycomb lattice (c) critical phase sphere in $p_x,p_y,p_z,p_I$ tetrahedra (d) phase diagram of radius vs $p_h$ in square lattice.}
    \label{fig:ph}
\end{figure}

Both of our models (eqs.~\ref{eq:YL},~\ref{eq:SqL}) inherit the entanglement phases of the original Kitaev model in the absence of single site or two-site interaction terms. 
They exhibit a critical phase when the probabilities $p_\alpha$ are similar to each other, with a transition to an area-law phase in the anisotropic limit with topologically protected qubits as shown in Fig.~\ref{fig:ph}, $p_h=0$ subfigures.
All phases have been found to be rotationally symmetric, circular for the honeycomb lattice or spherical for the square lattice, with respect to the center of the simplex (see Appendix~\ref{Appdx:spherical} for details). Therefore, we report the phase transition locations as distances from the isotropic point. This additionally allows us to compare quantities between the honeycomb and square lattice models. Concretely, we define $r = ||(p_x,p_y,p_z) - (\frac{1}{3},\frac{1}{3},\frac{1}{3})||$ for the honeycomb lattice  and $r = ||(p_x,p_y,p_z,p_I) - (\frac{1}{4},\frac{1}{4},\frac{1}{4},\frac{1}{4})||$ for the square lattice. For the following figures, we normalize these radii to the distance from the isotropic point to the edge-center of the enclosing simplex, $r_{edge}$. For the Kitaev model transition is at $(p_x,p_y,p_z)=(\frac{1}{6},\frac{1}{6},\frac{2}{3})$ which implies $r_c=\frac{1}{\sqrt{6}}$ and $r_c/r_{edge}=1$. This carries over for our honeycomb lattice model. The square lattice model without interaction terms, however, has with $r_c/r_{edge} = 0.96$.

The critical phase nearer the isotropic point is characterized by a logarithmic violation of the area-law with a leading term $S_A \sim L \ln L$. This scaling is reminiscent of gapless spin-liquid states with a Majorana Fermi surface, which also exhibit $L\log L$ entanglement entropy. 

The area-law phase is characterized by an entanglement entropy scaling of $S \approx \alpha L - S_{\rm topo}$, an area-law with a constant topological term. We find $S_{\rm topo} = 1$, consistent with a gapped $Z_2$ spin liquid. In the limit where one probability goes to 1 this reduces to the toric code. This phase supports dynamically stabilized logical qubits: under purification dynamics the remaining entropy saturates at $S=2$ for times that are exponentially long in the system size. These two logical qubits correspond to the two non-contractable loops around either dimension of the system. Although we discussed the different flavors of loops earlier, our qudit generalization does not support additional logical qubits compared to the Kitaev model. The loops are labeled by $\sigma$ DOF at their endpoints, so closed loops, as these non-contractable ones are, do not carry any flavor.

As discussed above, tripartite mutual information is the clearest entanglement measure for the differentiation of the various phases.
In the critical phase, the tripartite mutual information takes the value $I_3(A:B:C)=-1$ \cite{Lavasani_PRB2023,Zhu_PRR2024}. This arises from a single independent long string with support across $ABC$. Although there are extensively many strings that connect $A$ and $C$ while crossing $B$, they are not linearly independent for the purpose of mutual information: given any set of $n$ such strings, one can multiply the first string by suitable products of plaquette stabilizers to deform the remaining $n-1$ so that their support on $B$ becomes trivial. Consequently, only one string is shared nontrivially by $A$, $B$, and $C$, yielding $I_3=-1$, while the remaining bipartite contributions cancel \cite{Lavasani_PRB2023}. For the honeycomb lattice, a representative long string has the form $g_P = P^\sigma_i \big(\prod_{m\in\{i,j\}} Z^\tau_m\big) P^\sigma_j$ with $P \in \{X,Y,Z\}$. These lead to three flavors $\{g_X,g_Y,g_Z\}$, but they are not independent: $\{g_X,g_Y,g_Z\}$ can be rewritten as $\{g_X g_Z,\, g_Y g_Z,\, g_Z\}$, canceling the support of the first two and leaving a single independent operator shared by $ABC$.

Deep in the topological area-law phase, we find $I_2(A:C)=+2$. Here, the stabilizer group is generated by plaquettes together with short strings. On the honeycomb lattice in the anisotropic limit (e.g., $p_z \gg p_x,p_y$), the first $+1$ contribution to $I_2(A:C)$ can be traced to an effective Bell pair like stabilizer built from the product of all plaquettes that intersect $B$. This product produces a pair of parallel noncontractible loops with support separately in $A$ and $C$, yielding mutual information between $A$ and $C$ but no support in $B$. A second $+1$ arises from a stabilizer shared between $A$ and $C$ that combines rows of plaquettes in $B$ with the bond terms. In this limit, the stabilizer group is generated by the plaquettes together with roughly $L^2$ bonds of the form $P^\sigma_i P^\sigma_j Z^\tau_i Z^\tau_j$ with $P\in\{X,Y,Z\}$; subject to one global constraint among the $Z$-type bonds, yielding a residual entropy $S=2$. Within this group, there exists a stabilizer built from alternating rows of plaquettes in $B$ combined with the three flavors of $Z$-bonds that has trivial support on $B$ but cannot be factored into operators supported only in $A$ or only in $C$, as shown in Fig. \ref{Fig:4}. This gives an additional unit of mutual information between $A$ and $C$. Adding these up, in the area-law phase all other stabilizers are short-ranged (aside from the product of plaquettes discussed above), so in the four-cylinder geometry the only nonvanishing contribution to $I_3(A:B:C)$ comes from $I_2(A:C)$ and the shared effective Bell pair. Hence $I_3(A:B:C)=+2$ in the topologically ordered area-law phase. We discuss the corresponding arguments for the square lattice model in Appendix~\ref{Appdx:MI}.

\subsubsection{Single-site measurements}

Fig. \ref{fig:ph} shows the effect of the onsite $Z^\sigma$ term on the transition radius normalized to the edge of the simplex. In the honeycomb lattice model, for $p_h=0$ the transition begins with $r_c/r_{edge} = 1$, increases until $p_h\approx 0.18$ then decreases to $r_c/r_{edge} \approx 0.95$ by $p_h=0.25$ and stays there for all higher values of $p_h$. Note that since all probabilities must sum to 1, the size of the simplex decreases as $p_h$ increases, so the constant normalized radius still corresponds to a shrinking transition probability along a given direction.

In the square lattice model the transition radius, now defining a sphere inside a tetrahedron rather than a circle inside a triangle begins at $r_c/r_{edge} = 0.96$, increases until $p_h\approx 0.20$ then decreases until $p_h\approx 0.65$, where the critical region completely disappears.

In both cases the introduction of $p_h$ leads to first an increase then a decrease in the size of the critical region. This can be understood from the string fusion rule induced by $Z^\sigma$ described above. For small $p_h$ allowing the fusing of different flavor strings creates longer strings, favoring the critical phase. However, if too many operators are intra-site, $p_h\gtrapprox 0.2$ then the probability of choosing a site that already has the endpoints of multiple finite length strings decreases and the onsite term only creates length-zero strings.

Although single-site Pauli checks are “error-like”, within our plaquette-preserving protocol they stabilize, rather than degrade, the quantum information stored in the logical qubits. As discussed above, for large $p_h$ the onsite measurements project many nonlocal degrees of freedom into trivial (length-zero) strings, sharply suppressing long strings. In the Kitaev model or the toric code single-site checks cause phase flips in the plaquettes, changing the expectation value of the noncontractable loops and hence flipping the value of the logical qubits. In contrast, here single-site checks commute with all plaquettes, meaning they do not interfere with the logical qubits. 
The net effect is that the topological sectors remain intact and suppression of long strings leads to the shrinking (honeycomb lattice) or eliminating (square lattice) of the critical region, thereby preserving the encoded logical qubits for a large range of parameters. For the honeycomb model this is valid for any local checks in the sigma sector of the form $P_i^\sigma$ with $P\in\{X,Y,Z\}$, whereas for the square lattice we use $Z_i^\sigma$ to maintain plaquette conservation. 
The trends in Fig.~\ref{fig:ph} reflect this stabilization of the area-law in both models.

\begin{figure}
    \centering
    \includegraphics[width=0.9\linewidth]{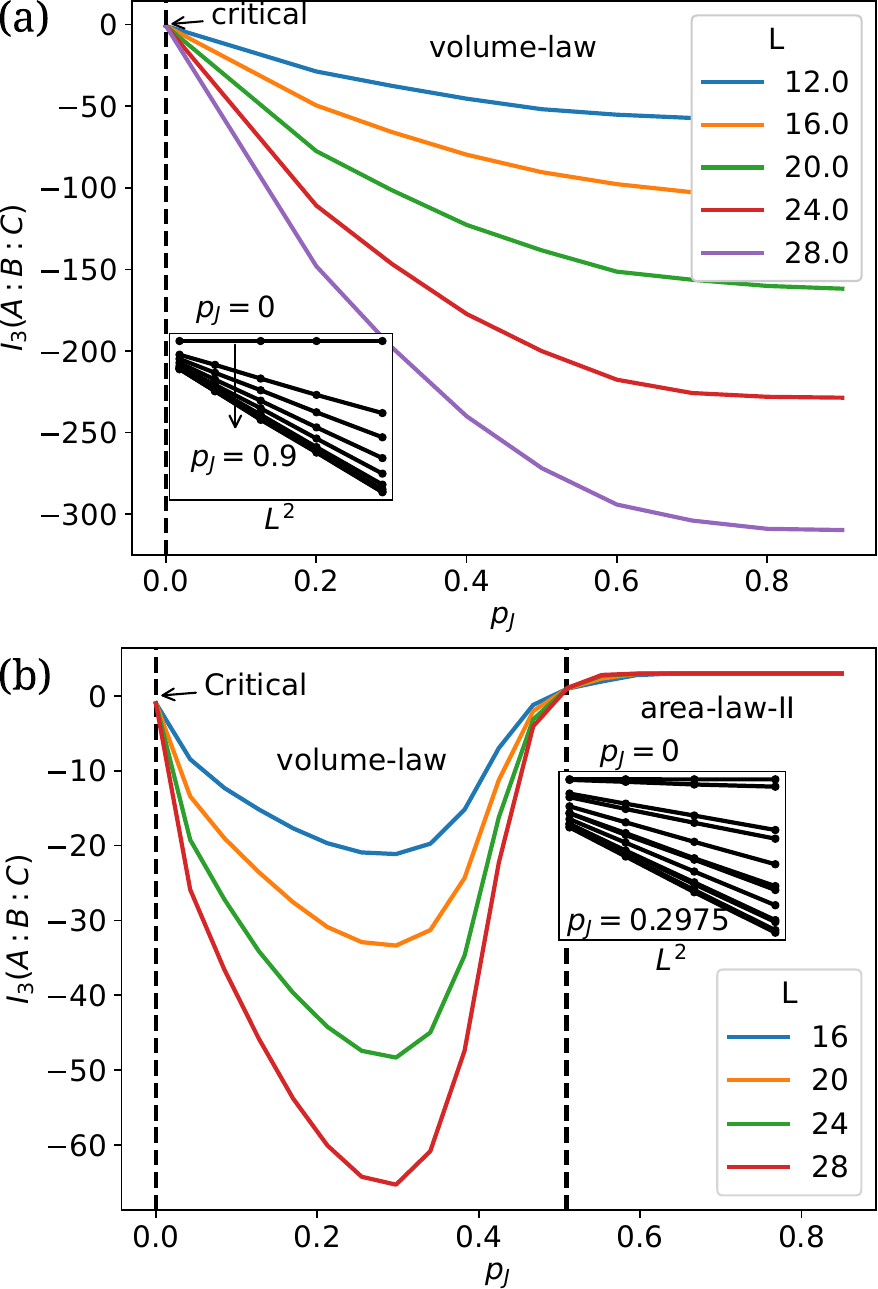}
    \caption{Cuts of varying $p_J$ keeping all other probabilities equal in the (a) honeycomb and (b) square lattices. Both show an immediate transition to the volume-law phase, and the square lattice shows a transition to a distinct area-law II phase above $p_J \approx 0.51$.}
    \label{fig:pj_cuts}
\end{figure}

\begin{figure}
    \centering
    \includegraphics[width=\linewidth]{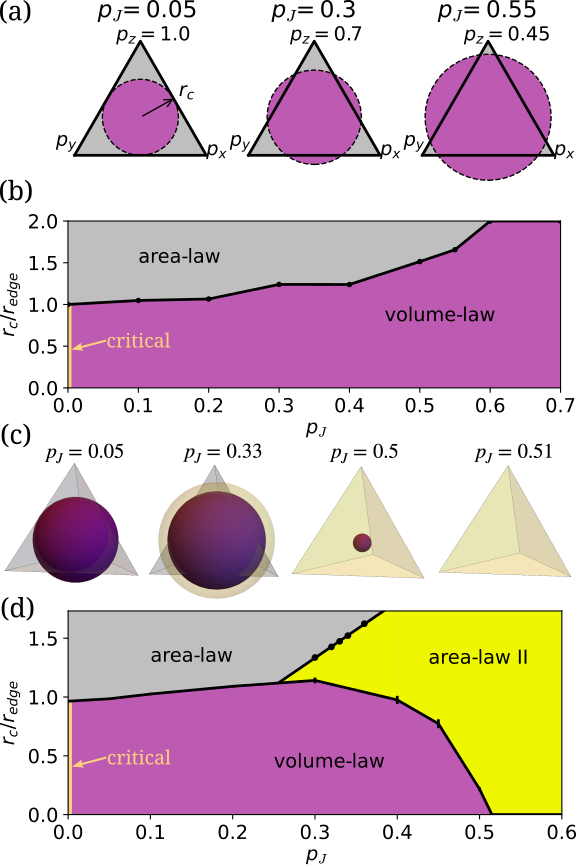}
    \caption{Effect of two-site term (a) area-law (gray) and volume-law (purple) in $p_x,p_y,p_z$ triangle for selected values of $p_J$ (b) phase diagram of radius vs $p_J$ in honeycomb lattice (c) volume-law and area-law-II spheres in $p_x,p_y,p_z,p_I$ tetrahedra (d) phase diagram of radius vs $p_J$ in square lattice showing new area-law-II phase (yellow).}
    \label{fig:pj_rc}
\end{figure}
\subsubsection{Two-site interaction measurements}
Fig. \ref{fig:pj_cuts} show cuts of the tripartite mutual information varying $p_J$ along the isotropic line where $p_x=p_y=p_z(=p_I)$. For the honeycomb lattice there is an immediate transition from the critical phase to the volume-law behavior for any $p_J$. The inset shows the $L^2$ scaling of the tripartite mutual information with a prefactor that increases with increasing $p_J$, saturating above $p_J\approx 0.8$. The square lattice shows the same immediate transition to a volume-law but an additional area-law phase appears above $p_J = 0.51$. The prefactor in the volume-law peaks just below $p_J\approx 0.3$ then decreases back to zero at the transition to area-law. The high $p_J$ area-law phase is not the same as the one present for $p_J=0$ and anisotropic probabilities. This new phase has $I_3=3$ and has three logical qubits, as seen in the dynamical purification. The repeated measurements of $Z^\sigma_iZ^\sigma_j$ suppress the measurement of $X^\sigma,Y^\sigma$ DOF as they anti-commute. 

The area-law–II regime can be understood in the large–$p_J$ limit of the square-lattice model. In this limit, the stabilizer group is generated by all plaquette operators together with the bond parities $Z^\sigma_i Z^\sigma_j$ (on any of the $X,Y,Z,$ or $I$ bonds). The code factorizes into a tensor product of (i) a topological code generated by the plaquettes, which contributes two logical qubits on the torus, and (ii) a length-$N$ repetition code generated by nearest-neighbor $Z^\sigma$ parities,
\[
\mathcal{S}_{\mathrm{rep}}=\{\,Z^\sigma_1 Z^\sigma_2,\; Z^\sigma_2 Z^\sigma_3,\; \dots,\; Z^\sigma_{N-1} Z^\sigma_N\,\},
\]
with one $Z^\sigma_i Z^\sigma_j$ per site. On a torus there is one global constraint among these parities, so for $N$ qubits there are $N-1$ independent stabilizers and the code space has dimension $2^N/2^{N-1}=2$, i.e., one logical qubit. This repetition-code logical is
\[
\mathrm{Span}\{\ket{0}^{\otimes N},\,\ket{1}^{\otimes N}\},
\]
with logical Pauli operators $\bar{X}=\prod_i X_i^\sigma$ and $\bar{Z}=Z^\sigma_1$ (equivalently, any single-site $Z^\sigma_i$). The code corrects up to $\lfloor (N-1)/2 \rfloor$ bit flips ($X$ errors), but a single phase flip ($Z$ error) implements the logical $\bar{Z}$, so the overall code distance is $1$. Combined with the two topological qubits, this yields three logical qubits in the area-law–II phase.
As the $p_J$ terms in the honeycomb lattice model allow measurement of any $X^\sigma_iX^\sigma_j,Y^\sigma_iY^\sigma_j,Z^\sigma_iZ^\sigma_j$ along a bond, there is no possibility of a global phase qubit emerging and the system stays in the volume-law.

Fig. \ref{fig:pj_rc} shows the phase diagram for varying $p_J$ and the normalized radius, as defined as above, for both lattices. In both cases there is an immediate transition to the volume-law phase. For the honeycomb lattice the size of the volume-law region expands until, near $p_J\approx 0.6$, it encompasses the entire region, with the system staying in the volume-law regime for any values of $p_x,p_y,p_z$ above that. The square lattice shows the volume-law to new area-law transition along the isotropic line, corresponding to $r_c=0$ in this diagram. For more anisotropic points, above $r\approx 1.14$, there is a direct transition from the AL-I to AL-II phase. For the region $0.26<p_J<0.38$ all three phases exist as concentric spheres when varying the remaining probabilities. The AL-II sphere increases in radius until it takes up the entire tetrahedra at $p_J\approx 0.38$ while the volume-law sphere decreases in radius until it disappears near $p_J\approx 0.51$.

\section{Conclusion}
We studied the measurement-only dynamics of the qudit generalizations of Kitaev model on honeycomb and square lattices. We showed that the dynamics can be mapped onto multiflavor loop models where certain single qubit errors fuse different flavors. Our simulations indicate that these errors can stabilize the area-law phase that hosts logical qubits. We also investigated certain two-qubit measurements that induce interactions among Majorana fermions. We showed that these terms can lead to volume-law phases for both models while a new area-law phase with an additional logical qubit for the square lattice model for large enough coupling. Interesting future directions include multiflavor loop model simulations for exploring criticality in these models.

\section{Acknowledgments}
We thank Nandini Trivedi and Ryan Buechele for fruitful discussions. AV and OE acknowledge support from NSF Award No. DMR-2234352. EDR acknowledge support from NSF Award No. DMR-2206987. We thank the ASU Research Computing Center for high-performance computing resources.

AV and EDR contributed equally to this work.

\appendix
\section{Qudit (\texorpdfstring{$d=4$}{d=4}) generalization of the Kitaev model}\label{Appdx:qudit}

We briefly detail the $d=4$ (qudit) generalizations of the Kitaev model on honeycomb and square lattices. The exact solvability of the spin-$\tfrac{1}{2}$ Kitaev model exploits the Pauli algebra, $\{\sigma^\alpha,\sigma^\beta\}=2\delta^{\alpha\beta}$, and the identity $\sigma^x\sigma^y\sigma^z=i$, i.e., a realization of the Clifford algebra in three generators. To extend this structure to a four-dimensional on-site Hilbert space, we employ a set of $4\times4$ Gamma matrices $\{\Gamma^a\}_{a=1}^{5}$ that furnish a representation of the five-dimensional Clifford algebra,
\[
\{\Gamma^a,\Gamma^b\}=2\delta^{ab}, \qquad (a,b=1,\dots,5).
\]
Together with the ten bivectors $\Gamma^{ab}=\tfrac{i}{2}[\Gamma^a,\Gamma^b]$ ($a<b$) and the identity, these matrices span the local Hilbert space on a $d=4$ qudit.

For honeycomb and square lattice, the Hamiltonian is
\begin{equation}
H \;=\; -\sum_{\langle ij\rangle_\alpha} K^\alpha \Big( \Gamma_i^\alpha \Gamma_j^\alpha \;+\; \sum_{\beta\in\mathcal{R}} \Gamma_i^{\alpha\beta} \Gamma_j^{\alpha\beta} \Big),
\label{eq:GammaHamiltonian}
\end{equation}
where $\alpha\in\{1,\dots,\alpha_m\}$ with $\alpha_m=3$ (honeycomb) or $4$ (square). The “residual” index set $\mathcal{R}$ collects the remaining Gamma components not used as bond labels: $\mathcal{R}=\{4,5\}$ for honeycomb and $\mathcal{R}=\{5\}$ for square. Equation~\eqref{eq:GammaHamiltonian} reduces to the standard Kitaev form when only the first term is retained, and the additional $\beta\in\mathcal{R}$ couplings implement the qudit generalization used here.

It is convenient to realize the $d=4$ on-site Hilbert space as a tensor product of two qubits, $\mathcal{H}_{\text{site}}\simeq \mathbb{C}^2_\sigma \otimes \mathbb{C}^2_\tau$, with Pauli operators $\sigma^\mu$ and $\tau^\mu$ ($\mu=x,y,z$). A concrete representation of the Gamma matrices is
\begin{equation}
\Gamma^\alpha=-\sigma^y\otimes\tau^\alpha,\quad
\Gamma^4=\sigma^x\otimes\mathbb{I},\quad
\Gamma^5=\sigma^z\otimes\mathbb{I}.
\label{eq:GammaSpinOrbital}
\end{equation}
where $\alpha=x,y,z$.
Substituting \eqref{eq:GammaSpinOrbital} into \eqref{eq:GammaHamiltonian} yields the spin–orbital (``$\sigma$–$\tau$'') forms used in the main text,\ Eqs.~(\ref{eq:YL}) and~(\ref{eq:SqL}).

For exact solvability, it is useful to fractionalize the Gamma operators into Majoranas. We introduce six Majorana fermions per site,
\[
\Gamma_i^a = i\, b_i^a\, c_i,\qquad
\Gamma_i^{ab} = -i\, b_i^a b_i^b \quad (a<b),
\]
with $a\in\{x,y,z,4,5\}$, one itinerant Majorana $c_i$, and five gauge Majoranas $b_i^a$. The physical Hilbert space is obtained by imposing the local gauge constraint
\[
D_i \equiv b_i^x b_i^y b_i^z b_i^4 b_i^5 c_i = +1,
\]
which projects from the enlarged Majorana Fock space to the qudit subspace. In this language, the plaquette fluxes and bond bilinears appearing in \eqref{eq:GammaHamiltonian} map to commuting $\mathbb{Z}_2$ gauge fields coupled to the itinerant $c$ Majoranas, yielding the free-Majorana equations quoted in Eqs.~(\ref{eq:MFYL}) and~(\ref{eq:MFSqL}).

\section{Mutual Information for the square lattice model} \label{Appdx:MI}

In the critical regime an analogous argument to the honeycomb lattice holds on the square lattice too. Here the long string takes the form $g_P = P^\sigma_i \big(\prod_{m\in\{i,j\}} Z^\sigma_m Z^\tau_m\big) P^\sigma_j$ with $P \in \{X,Y\}$; again the two flavors reduce to $\{g_X g_Y,\, g_Y\}$, so $I_3(A:B:C)=-1$ throughout the critical regime.

For the topological area-law, a similar but slightly more involved construction than the honeycomb lattice applies to the square lattice. In addition to the effective Bell pair like stabilizer shared between regions $A$ and $C$, we work in an anisotropic limit with $p_x>p_y\gg p_z,p_I$. There are two bond types, $X$ and $Y$, each with two flavors in the doubled $(\sigma,\tau)$ sectors. Because the number of plaquettes is $L^2-1$ while the two-flavor bond set contains $2L^2-1$ independent generators, the stabilizer tableau can be completed using only one flavor per bond type.
A convenient choice takes alternating rows of plaquettes together with $X^\sigma_i X^\sigma_j X^\tau_i X^\tau_j$ on the $X$ bonds and $Y^\sigma_i Y^\sigma_j Y^\tau_i Y^\tau_j$ on the $Y$ bonds. Equivalently, one may use $X^\sigma_i X^\sigma_j Y^\tau_i Y^\tau_j$ on $X$ bonds and $Y^\sigma_i Y^\sigma_j X^\tau_i X^\tau_j$ on $Y$ bonds. The product of these bond terms with the selected plaquettes yields a stabilizer that has trivial support on $B$ but cannot be written with support only in $A$ or only in $C$ as shown in Fig. \ref{Fig:4}. By the same reasoning as in the honeycomb case, this construction leads to $I_3(A:B:C)=+2$ in the area-law regime.

\section{Spherical symmetry of the phase diagram} \label{Appdx:spherical}
For the phase diagram scans at constant single-site measurement rate $p_h$ or alternatively two-site measurement $p_J$, we re-parameterize the remaining probabilities $(p_x, p_y, p_z, p_I)$ by spherical coordinates $(r, \theta, \phi)$ in a tetrahedron frame. The construction keeps the normalization
$p_x + p_y + p_z + p_I + p_h = 1$
by writing the four barycentric coordinates as a uniform share plus a directional deviation:
$$p_\alpha = \frac{1 - p_h}{4} + ru_\alpha(\theta,\phi), \qquad \alpha \in {x,y,z,i}$$.

The directional weights $u_\alpha(\theta,\phi)$ form a traceless 4-vector $\sum_\alpha u_\alpha = 0$ aligned with the tetrahedral symmetry axes. In the implementation,
\begin{align*}
    u_x(\theta,\phi) &= \frac{\sqrt{3}}{4}\big[\sin\theta(\cos\phi - \sin\phi) - \cos\theta\big],\\
u_y(\theta,\phi) &= \frac{\sqrt{3}}{4}\big[\sin\theta(\sin\phi - \cos\phi) - \cos\theta\big],\\
u_z(\theta,\phi) &= \frac{\sqrt{3}}{4}\big[-\sin\theta(\cos\phi + \sin\phi) + \cos\theta\big],\\
u_I(\theta,\phi) &= \frac{\sqrt{3}}{4}\big[\sin\theta(\cos\phi + \sin\phi) + \cos\theta\big].
\end{align*}

At $r = 0$ one sits at the tetrahedron center with $p_x = p_y = p_z = p_I = (1 - p_h)/4$. Positivity of all probabilities bounds the allowed radius along $(\theta,\phi)$ by
$r_{\max}(\theta,\phi; p_h) = \min_{\alpha:,u_\alpha(\theta,\phi) < 0} \frac{(1 - p_h)/4}{|u_\alpha(\theta,\phi)|}$,
the first contact with a tetrahedron face where one $p_\alpha$ vanishes.

\begin{figure}
    \centering
    \includegraphics[width=0.8\linewidth]{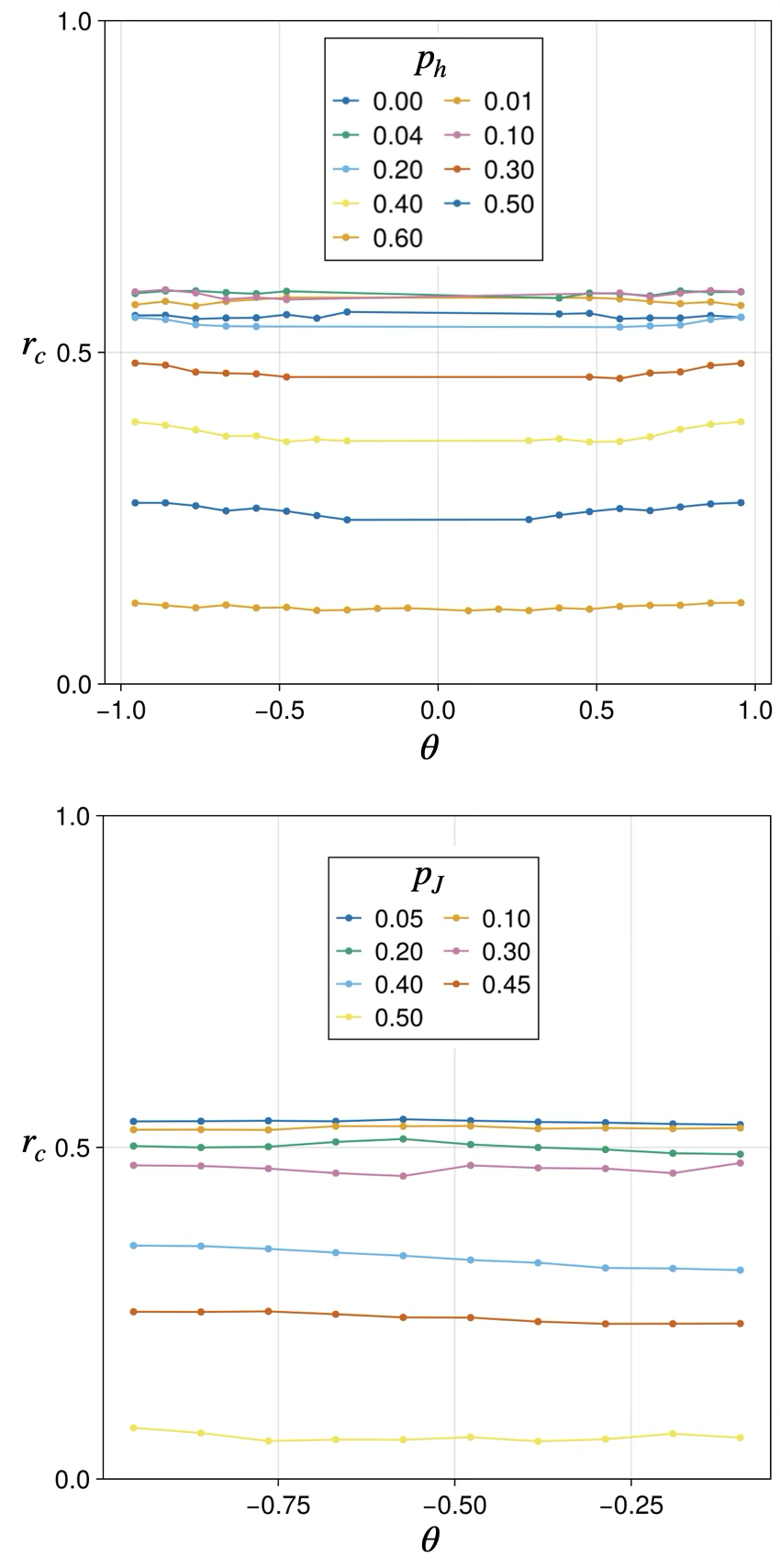}
    \caption{Top: Critical radii $r_c$ of the constant $p_h$ phase diagrams with varying polar coordinate $\theta$ for $\phi=-2.35$. Bottom: Same analysis for the constant $p_J$ phase diagrams.}
    \label{fig:A1}
\end{figure}
To extract the critical surface, we fix $(p_h, \phi)$ and scan $r$ from the center toward the boundary along a given $\theta$. For each $(r,\theta,\phi; p_h)$ we convert $(r,\theta,\phi) \rightarrow (p_x,p_y,p_z,p_i)$ using the mapping above, evolve the monitored circuit, and compute the entanglement measures (specifically tripartite information $I_3$) averaged over 100 trajectories. Using the same criterion described in Sec. III B, we define $r_c(\theta; \phi, p_h)$ as the smallest r at which the curves cross from the critical behavior (interior of the sphere) to the area-law behavior. The scan respects the geometric bound $r \le r_{\max}(\theta,\phi; p_h)$.

Fig. \ref{fig:A1} shows critical radius $r_c$ versus polar angle $\theta$ at fixed $(p_{h(J)}, \phi)$. Within numerical precision, $r_c$ is independent of $\theta$, indicating a spherical phase boundary centered at $(p_x,p_y,p_z,p_i) = \big((1 - p_h)/4,(1 - p_h)/4,(1 - p_h)/4,(1 - p_h)/4\big)$. We have similarly checked that $r_c$ does not depend on the azimuthal angle $\phi$ at fixed $p_h$, consistent with spherical symmetry.

We repeat the same analysis for the two-site interaction terms and obtain similar spherical phase boundaries as depicted in Fig \ref{fig:A1}.

\begin{figure}[t!] 
\includegraphics[width=0.8\linewidth]{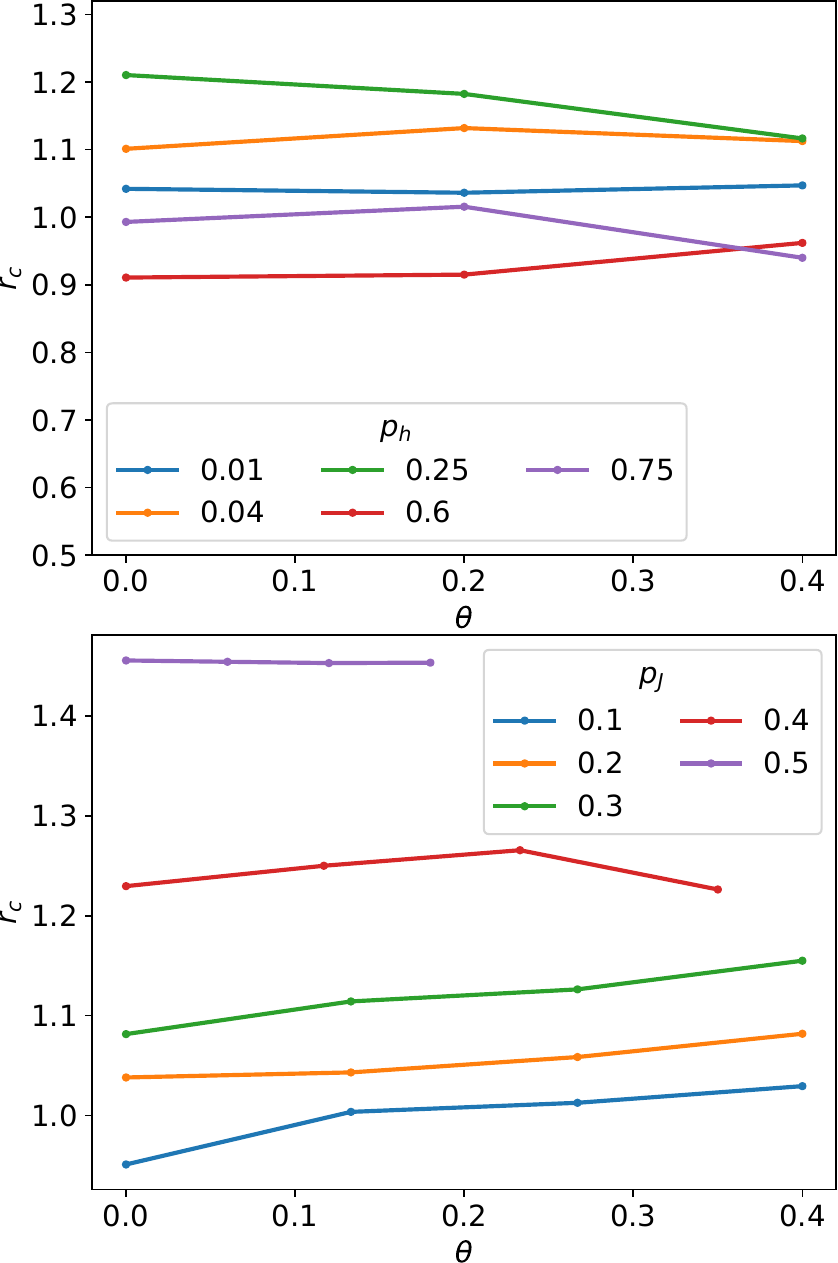}
    \caption{Top: Critical radii $r_c$ at constant $p_h$ varying the angle $\theta$ for the honeycomb lattice. Bottom: Same analysis for constant $p_J$ cuts.}
    \label{fig:A2}
\end{figure}

A similar, though somewhat noisier trend holds for the honeycomb lattice, shown in Fig. \ref{fig:A2}. Here, as there are only three parameters and one constraint, only a single angle is necessary. Note that for the two-site interaction the angular range of the volume-law region inside the triangle of allowable parameter combinations becomes smaller as the critical radius increases. For $p_J$ above $0.5$ a single cut along the $p_x=p_y$ line was used to determine the critical radius.

\bibliography{references.bib}
\end{document}